\documentclass{iaus}
\usepackage{graphicx}

\title[MHD Remote Numerical Simulations] 
{MHD Remote Numerical Simulations: Evolution of Coronal Mass Ejections}

\author[Hern\'andez--Cervantes, \etal]   
{L. Hern\'andez--Cervantes$^1$%
  \thanks{Present address: Instituto de Astronom\'\i a, UNAM, 04510, Mexico
City, Mexico},
 A. Santill\'an$^2$ \break \and A.R. Gonz\'alez--Ponce$^3$}

\affiliation{$^1$ Instituto de Astronom\'\i a, UNAM, 04510, Mexico City, 
Mexico \break email: liliana@astrosu.unam.mx\\[\affilskip]
$^2$Direcci\'on General de Servicios de C\'omputo Acad\'emico, UNAM, 04510,
Mexico City, Mexico\\[\affilskip]
$^3$ Instituto de Ecolog\'\i a, UNAM, 04510, Mexico City, Mexico}

\pubyear{2009}
\volume{259}  
\pagerange{119--126}
\date{?? and in revised form ??}
\jname{Cosmic Magnetic Fields: From Planets, to Stars and Galaxies}
\editors{K.G. Strassmeier, A.G. Kosovichev \& J. Beckmann, eds.}
\begin{document}

\maketitle

\begin{abstract}
Coronal mass ejections (CMEs) are solar eruptions into interplanetary space of 
as much as a few billion tons of plasma, with embedded magnetic fields from the 
Sun's corona. These perturbations play a very important role in 
solar--terrestrial relations, in particular in the spaceweather. In this work we
present some preliminary results of the software development at the Universidad
Nacional Aut\'onoma de M\'exico to performe Remote MHD Numerical Simulations.
This is done to study the evolution of the CMEs in the interplanetary medium 
through a Web--based interface and the results are store into a database. The 
new astrophysical computational tool is called the Mexican Virtual Solar 
Observatory (MVSO) and is aimed to create theoretical models that may be helpful
in the interpretation of observational solar data.


\end{abstract}

\firstsection 
\section{Introduction}

The Mexican Virtual Solar Observatory (MVSO) is a set of software tools that 
offer global solutions for Web development. The operating system is Linux, the 
Web server is Apache, SQL (Structure Query Language) and the relational 
database management system is MySQL, everything is programmed with PHP 
(Hypertext Pre--Processor). The computational backbone of the MVSO is 
structured into three stages. The first part is the related to the graphic user 
interface (GUI), the second part is associated to the remote numerical 
simulations (RNS) and the third part is the creation of the database and 
associated search tools. The implementation is explained by \cite{Hernandez2008}

\section{Results}

\begin{figure}
\includegraphics[height=3in,width=5.3in]{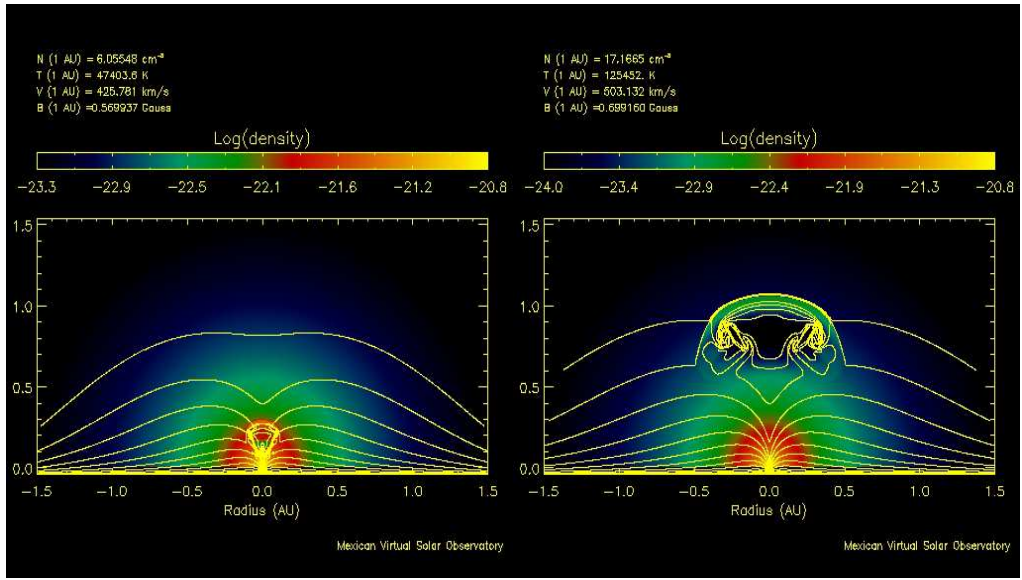}
  \caption{
Evolution of the CME in the magnetized solar wind. The figure show the
density (\textit{color logarithmic scale}) and intensity of the total magnetic
field (\textit{solid lines}) at two select times: 6 and 78 hours after the 
inyection of the perturbation.}\label{fig:mvso}
\end{figure}

All calculations of the evolution of the CME in the magnetized solar wind are 
performed with the MHD code ZEUS-3D, which solves the 3D 
system of ideal MHD equations by finite differences on fixed Eulerian mesh 
(\cite{StoneNorman1992a}). The MVSO 
uses a simplified model to understand the dynamics of a CME in the 
interplanetary space (\cite{Santillan2008}). Initially, we produce the ambient 
solar wind by specifying the fluid velocity, magnetic field, density, and 
temperature at an inner boundary of the grid, which is located beyond the 
critical point (r = 18 Ro $\sim$ 0.083 AU), and then the wind is allowed to 
evolve and reach a stationary equilibrium. For the injection of the magnetic 
field we used the technique described by \cite{StoneNorman1992b}; this consists 
of using time dependent analytic solutions of the \cite{Low1984} models. 
Finally we add an ejecta--like perturbation at the inner boundary to simulate 
the appearance of the CME into de interplanetary medium.
Typical results produced by the MVSO are shown in the two snapshots displayed 
in the figure~\ref{fig:mvso}, where the density is shown in logarithmic 
color--scale along with the intensity of the total magnetic field 
(\textit{solid lines}). 
The rigidity and elasticity given to the solar wind by the magnetic field is 
better accentuated in 2D when the plane of motion of the CME is 
parallel to the field lines and the colliding gas distorts the initial field 
configuration. We illustrate the response of these deformed field lines in the 
two snapshots displayed in figure~\ref{fig:mvso}. The tension of the magnetic 
field dominates the evolution and the results are completely different from 
those of the purely hydrodynamic cases. The physical quantities ($n$, $T$, 
$\bf v$ \& $\bf B$) of the medium at 1 AU change drastically, when the 
disturbance crosses by this point.  For example, the density increase a factor 
$\sim$ 3 and the size of the perturbed region has grown close to 1 AU only 78 
hours after the inyection of the CME. This is clearly seen in the last snapshot 
of the figure~\ref{fig:mvso}.

\begin{acknowledgments}
We are grateful to Pepe Franco for useful comments. This work has been 
partially supported from DGAPA-UNAM grant IN104306 and CONACyT proyect 
CB2006--60526.
\end{acknowledgments}

\end{document}